\documentclass[conference]{IEEEtran}

\usepackage{tikz}
\usepackage{amsmath}

\usepackage{filecontents}
\usepackage{cite}
\usepackage{amsmath,amssymb,amsfonts}
\usepackage{graphicx}
\usepackage{textcomp}
\usepackage{tabularx}
\usepackage{subcaption}
\usepackage{makecell}
\usepackage{url}
\usepackage[switch]{lineno}
\usepackage{listings}
\usepackage{multirow}

\usepackage[margin=0.5in]{geometry}
\usepackage{textcomp}

\usepackage{listings}

\colorlet{punct}{red!60!black}
\definecolor{background}{HTML}{EEEEEE}
\definecolor{delim}{RGB}{20,105,176}
\colorlet{numb}{magenta!60!black}

\begin{document}

\date{}

\title{Preprint: Open Source  Compiling for V1Model RMT Switch: Making Data Center Networking Innovation Accessible
(Accepted for publication in 15th IEEE/ACM International Conference on Utility and Cloud Computing (UCC2022))}

\author{
{\rm Debobroto Das Robin,  Javed I. Khan}\\
Kent State University
} 

\maketitle
\begin{abstract}

  Very few of the innovations in deep networking have seen data center scale implementation.
  Because, the Data Center network's extreme scale performance requires hardware 
  implementation, which is only accessible to a few.   However, the emergence of 
  reconfigurable match-action table (RMT) paradigm-based switches 
  have finally opened up the development life cycle of data plane devices. 
  The P4 language is the dominant language choice for programming these devices. 
  Now, Network operators can implement the desired feature over white box RMT switches. 
  The process involves an innovator writing new algorithms in the P4 language and getting them compiled 
  for the target hardware. However, there is still a roadblock. 
  After designing an algorithm, the P4 program's compilation technology is not fully open-source. 
  Thus, it is very difficult for an average researcher to get deep insight into the performance 
  of his/her innovation when executed at the silicon level. There is no open-source compiler backend 
  available for this purpose. Proprietary compiler backends provided by different hardware vendors are available for this purpose. 
  However, they are closed-source and do not provide access to the internal mapping mechanisms. 
  Which inhibits experimenting with new mapping algorithms and innovative instruction sets for  
  \textit{reconfigurable match-action table}  architecture.
  This paper describes our work toward an open-source  compiler backend for compiling 
  P4\textsubscript{16} targeted for the V1Model architecture-based programmable switches.


  \end{abstract}

\maketitle

\section{Introduction}

In recent years, multistage reconfigurable match-action table (RMT)~\cite{bosshart2013forwarding} based  pipeline   
have become  the dominant architecture for programmable switches. 
They have seen widespread deployment in cloud data centers for their capability to bring programmability in the network data plane. 
The P4 language has emerged as the de-facto standard language to program these switches.


\textbf{A P4 compiler for RMT switches}: The P4 language~\cite{p416}   provides high-level imperative  
constructs to express packet processing logics for various packet processing architectures.  
Hence, there is no direct mapping between the  P4 program
and the RMT architecture's components. A P4 compiler  is required~\cite{bosshart2014p4} 
to translate a given P4 program into a target-specific 
executable program (hardware configuration binary) to be executed by a \textit{target switch}. 
The compiler frontend and midend are responsible~\cite{budiu2016architecture} for hardware independent syntax analysis and generating. 
a target-independent intermediate representation (\textit{IR}) of the P4 program. 
The \textit{\textbf{backend}} is target-dependent and it is responsible
for  mapping the IR components on the target
hardware resources. It works in two phases: 
a) \textit{\textbf{mapping-phase}}: here, the compiler backend
computes~\cite{P4StudioStructure} the P4 program's \textit{header field} to the RMT hardware's packet header vector 
(PHV) mapping (a.k.a. header mapping),
packet header parser state machine (represented as \textit{Parse Graph} in the IR) to RMT hardware's 
parser state table mapping (a.k.a. parse graph mapping) and 
the P4 program's control flow (represented as a graph of logical match-action tables) to RMT hardware's physical match-action 
table mapping (a.k.a. TDG mapping).
b) \textit{\textbf{executable binary generation phase}}: if the P4 program can be successfully mapped onto the
target hardware; corresponding \textit{hardware configurations} are generated from the 
mapping 
in the form of an executable hardware configuration binary.
This executable configuration is loaded into the target hardware
by the control plane and executed by the target hardware.

\textbf{Open source P4 compilers for RMT switch}: 
P4C~\cite{P4C} is the reference open-source compiler implementation for the P4 language. 
It supports two different RMT architecture based switches: 
a) simple\_switch model widely known as V1Model architecture~\cite{V1Model}
and b) Portable Switch Architecture (PSA)~\cite{PSA} developed by the  \textit{P4 Language Consortium} (not fully implemented yet). 
However, P4C  does not provide any  compiler backend for the real life target hardware of these two architectures. 
The P4C frontend+midend emits the intermediate representation, and the 
reference software switch implementation (BMV2~\cite{BMV2})  executes them over a 
CPU simulated version of the respective hardware architecture. 
It does not consider the architectural constraints
and practical hardware resource limitations~\cite{jose2015compiling} that exist in real target switches.
Hence, P4C can not decide about the realizability of the given P4 program over a specific instance of these RMT switches. 
Besides the P4C, several other open-source compilers for RMT architecture-based switches are available in the literature. 
However, some of them~\cite{jose2015compiling} only computes \textit{TDG mapping} and works
 with the older version (P4\textsubscript{14}~\cite{p414}) of the
P4 language, which is architecturally different from the current version of P4 (P4\textsubscript{16}~\cite{p416}). 
Some other works focus 
on different packet processing languages (e.g., Domino~\cite{sivaraman2016packet}),
or different hardware platforms (e.g., FPGA~\cite{wang2017p4fpga}). 
As a result, researchers need to use 
proprietary  compiler backends~\cite{P4Studio} to decide whether a P4 program can be implemented using an RMT switch or not. 
However, these systems are closed source, expensive, and often come with additional non-disclosure agreements~\cite{opentofino}.

\textbf{Why open-source compiler backend}: 
The compiler backend plays a crucial role in the P4 ecosystem by mapping a P4 program to the 
target hardware. It is responsible for measuring a P4 program's resource consumption in the RMT pipeline. 
Programmable switches contain a limited amount of hardware resources. Therefore, the P4 program with the least hardware resource 
required to achieve a specific task is more resource-efficient. 
In recent times, a large number of research works have 
used the BMV2~\cite{BMV2} simulator with the P4C compiler as their target platform, which lacks a 
compiler backend that can consider the real-life resource constraints present in the  target hardware. 
Without such a compiler backend, researchers can not measure the resource 
requirement of their schemes and can not compare the 
resource usage efficiency of multiple schemes. In the worst case, their P4 program may not 
be realizable at line-rate~\cite{sivaraman2016packet} using RMT switches, 
which are not even identifiable without a compiler backend. 
Thus, it stands to argue whether these P4 programs are directly executable over real-life RMT switches
without significant overhead.

A compiler backend needs to 
address several computationally intractable problems~\cite{jose2015compiling,vass2020compiling}
to find the mapping of a P4 program to the target hardware. 
The optimal algorithms often require a long time to finish~\cite{jose2015compiling,vass2020compiling}. 
With the growing rise of the \textit{in-network computing}~\cite{sapio2017network} paradigm,
various research works~\cite{robin2022clb,hauser2021survey,robin2022p4te,robin2021p4kp} are 
also focusing on delegating different 
\textit{network function}s in the  data plane. In these cases, researchers  
do not need to fit large P4 programs required for full-fledged switches with various features. 
Optimal mapping algorithms are useful when the data plane
programmers need to fit such large P4 programs into target
hardware. On the other hand, an open-source compiler backend that uses heuristic-based algorithms 
can  provide the researchers a quick decision about a 
smaller P4 program's realizability using a target hardware.
Which in turn can help the researchers to design improved network algorighms. 

The mapping algorithms used in the compiler backend are 
sensitive~\cite{jose2015compiling} to the resource (TCAM/SRAM storage, number of ALUs, crossbar width, etc.) 
requirements of a P4 program and 
the available resources in a target switch. 
The resource requirements of a P4 program can change at run time (e.g., an increase in the size of an IPv4 forwarding table), 
which can invalidate a previously computed mapping. 
With the rapid proliferation of \textit{network virtualization}~\cite{hancock2016hyper4} and 
\textit{Network-as-a-Service}~\cite{zhou2010services} 
paradigm, the requirement for on-demand network function deployment is also growing rapidly. 
It requires  quick and automated deployment of customized data plane algorithms on a short notice.
Therefore, developing faster and more efficient heuristic/approximate mapping
algorithms carry enormous significance here. 
With a closed source compiler backend, researchers can not experiment with different mapping algorithms. 
Besides this, there is a growing focus~\cite{hauser2021survey} on developing 
hardware units for supporting complex instructions (\textit{extern}~\cite{p416} in P4 language) in RMT architecture. 
Without an open-source compiler backend, researchers can not integrate newly developed externs in a P4 program and test their 
effectiveness.  
Independently developing compiler backend from scratch requires various common and repetitive tasks (i.e., IR parsing, 
representing parsed IR using a graph data structure, modeling hardware resources, etc.) not directly related to 
the computation of the mappings. An open-source compiler backend can allow researchers to 
focus on developing efficient mapping algorithms rather than focusing on the repetitive tasks.


Inspired by these factors, we focus on the following question: \textit{given that the IR for a P4 program is generated, 
how to bring innovation in IR-to-RMT hardware resource mapping?} By that, we focus on 
the design of an open-source P4 compiler backend (mapping phase only)
for \textbf{V1Model}~\cite{V1Model} architecture-based RMT switches. 

To the best of our knowledge, this is the first complete open-source P4\textsubscript{16} compiler backend that computes 
all three types of mapping for RMT architecture based programmable switches. 
The compiler backend requires two inputs: a) a specification of the available resources in a V1Model switch and 
b) the intermediate representation (IR) of a P4 program generated by the P4C frontend. As the P4C  does not provide 
any interface to specify the hardware resources of a V1Model switch, 
we have developed JSON format based hardware specification language (HSL) (sec.~\ref{HSLSection}) to express
the \textit{hardware resource specification}s of a V1Model switch. 
We briefly discussed 
the V1Model architecture in sec.~\ref{V1ModelArchitecture} along with the HSL (sec.~\ref{HSLSection}).
Then we present the structure of the IR provided by the P4C compiler frontend (sec.~\ref{IR}).
This backend uses various existing heuristic-based algorithms to allocate resources in the V1Model switch pipeline and 
computes  the \textit{IR to hardware} resource mapping. 
Once the mapping is found, computing the hardware configuration binaries 
requires a straightforward translation of  the mapping into hardware instruction  codes. 
As this work does not focus on 
executing a P4 program on any specific instance of V1Model switch, 
we leave the hardware configuration binary generation for future work. 
We discuss  the implementation and evaluation of our compiler backend in sec.~\ref{ImplementationAndEvaluation}. 
Then briefly discuss how it can be extended in sec.~\ref{Discussion}
and conclude the paper in sec.~\ref{Conclusion}.

\textbf{Why V1Model}: 
V1Model is the only RMT architecture fully supported by the open-source P4C compiler frontend,
used as a reference hardware architechture by large number of research works~\cite{hauser2021survey}, and  all
major programmable switch hardware vendors~\cite{opentofino}. 
Due to these reasons, we have chosen to 
build the compiler backend for this architecture.

\section{V1Model Architecture  } \label{V1ModelArchitecture} 
Here, we briefly discuss different components (along with their resources) of V1Model architecture and 
present a hardware 
specification language to represent a V1Model switch's resources. 


\begin{figure}[b]
 \centering
 \includegraphics[trim=0in 1in 0in 0, clip,scale=.345]{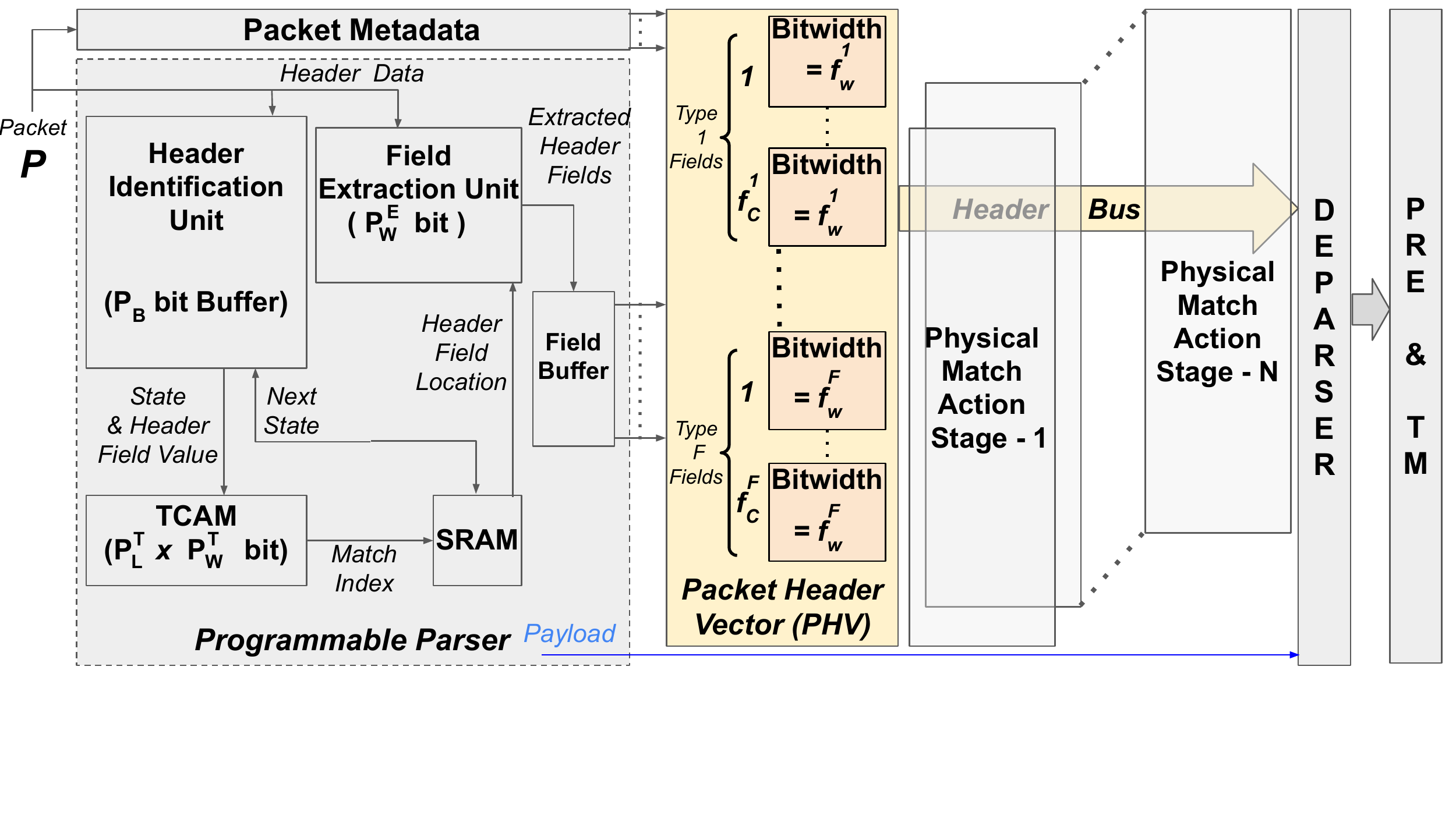}
 \caption{ V1Model pipeline architecture}
 \label{fig:V1ModelArchitecture}
\end{figure}

\subsection{Parser and Packet Header Vector} 
In V1Model architecture, an incoming packet at first goes through a TCAM based~\cite{gibb2013design} 
\textit{programmable parser} (fig.~\ref{fig:RMTSingleStage}), which executes the parsing logic provided 
in the form of a  state machine (converted to a \textit{state table} 
by a compiler backend).  The \textit{Header Identification Unit} of the parser 
contains a $P_B$ bit wide buffer to look ahead in the packet and identify maximum $H$ headers every cycle. 
It also contains a  
TCAM capable of storing $P_L^{T}$ entries to implement the parsing logic provided as \textit{state table}. 
At every cycle, maximum $f_C^T$ lookup field values (each having maximum lookup width $f_W^T$ b)   
and the \textit{current state} can be looked up in the TCAM. 
Based on the matching result the \textit{extraction unit} extracts maximum $P_W^{E}$ bit wide data as header fields 
and store them in a \textit{field buffer}. 
At every cycle, a few header fields are extracted, and 
the \textit{next parsing state} is fed to the \textit{header identification} unit for matching in the  TCAM in the next cycle.
After completing the parsing, all the extracted header fields  are sent to 
a \textit{Packet Header Vector} (PHV) from the \textit{field buffer}. 
The PHV is passed to subsequent components ($N$ match-action stages fig.~\ref{fig:V1ModelArchitecture}) 
in the pipeline through a wide \textit{header bus}.

\subsection{Match-Action Stages} \label{MatchActionStagesSubSection}

\begin{figure}[b]
 \centering
 \includegraphics[trim=0.1in 0in 0in 0, clip,scale=.35]{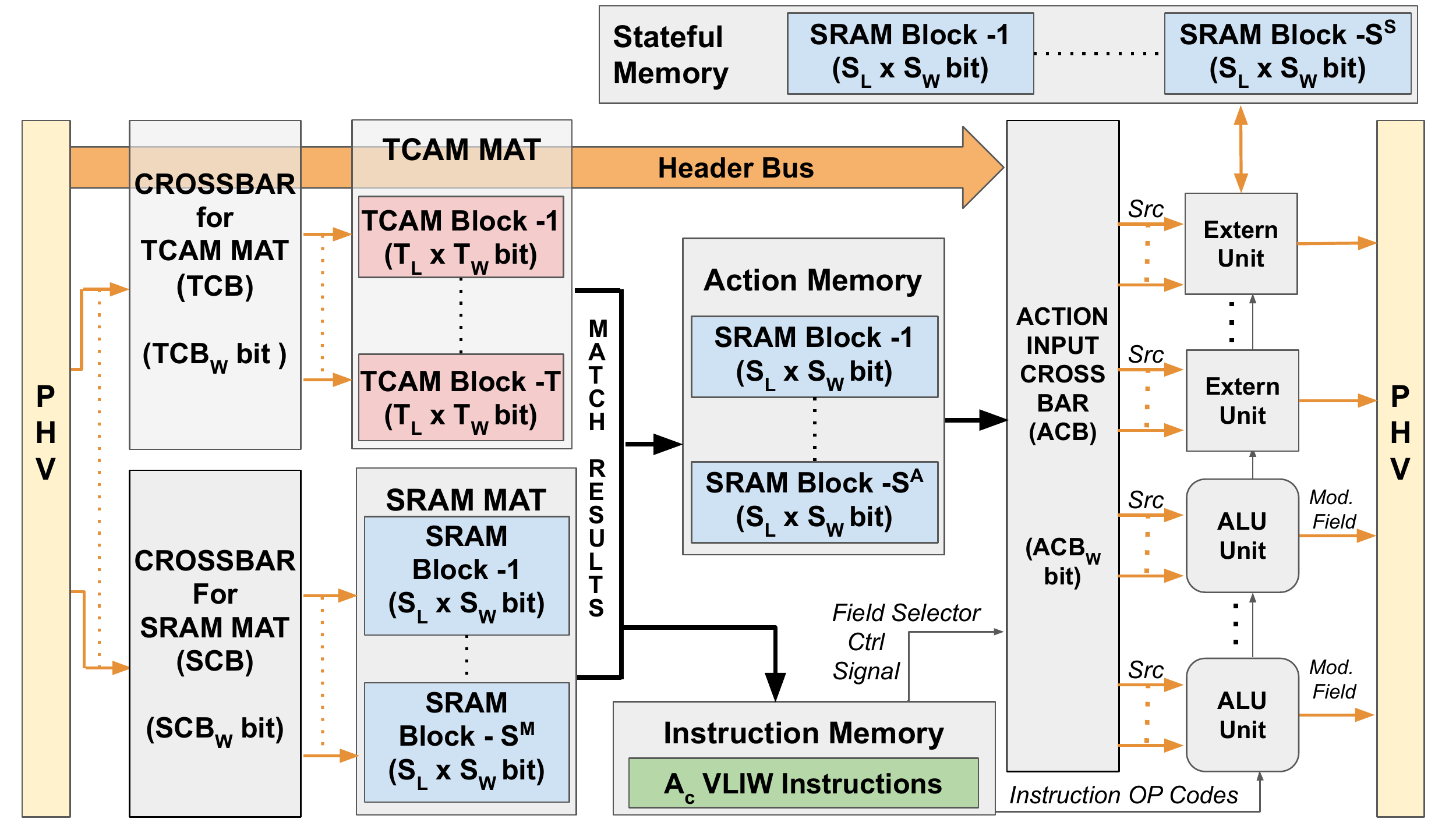}
 \caption{ A match-action stage in RMT pipeline}
 \label{fig:RMTSingleStage}
\end{figure}

Next, the PHV goes through a series of $N$ match-action stages for \textit{ingress stage} processing. 
Each stage (fig.~\ref{fig:RMTSingleStage}) contains $T$ units of $T_{W}$ bit wide  TCAM blocks and $S$ units of $S_{W}$ bit wide SRAM blocks
 each capable of storing  $T_{L}$ and $S_{L}$ entries respectively. 
The TCAM blocks are used to implement   \textit{physical match-action table}s (MAT) for ternary/range/prefix/exact matching. 
A fraction of the SRAM blocks ($S^M$ blocks) 
are used to implement hashtable (using $HS_K$ way Cuckoo hashtable~\cite{bosshart2013forwarding})  
based \textit{physical match-action table}s   
for exact matching, and the rest 
are used for storing other information (i.e., action arguments, next MAT address, etc.). 
These smaller \textit{physical match-action tables} can be run independently or  grouped 
together to match wider header fields within a stage or longer table across multiple stages. 
Header fields are supplied from PHV to the TCAM and SRAM based physical MATs through two crossbars, TCB (${TCB}_W$ bit wide) 
and SCB (${SCB}_W$  bit wide), respectively. 
With every entry in the MATs, there is a pointer to corresponding action information (action arguments, action instruction, 
address of the next MAT to be executed, etc.).  On finding a match in the MATs, the corresponding action 
information is loaded from the memory. Every match-action stage contains a separate arithmetic logic  unit (ALU) for every field of the PHV 
and for a fixed number of \textit{extern} units (hash, counter, register, meter, etc.)  for computation.

Every stage can store ${A}_{C}$ VLIW instructions for all the physical MATs. Every VLIW instruction carries separate instruction for  
the per header field ALU and extern units. Data is provided to these processing units from PHV through 
an ${ACB}_W$ bit wide crossbar (ACB). 
The action information (except the action instructions are stored in a 
dedicated memory) and the stateful memories used by the \textit{extern} units  
are allocated in separate chunks of the available SRAM blocks ($S^A$ blocks and $S^S$ SRAM blocks, respectively). 
Every stage contains $M_{P}$   memory ports (each one $M_{BW}$ bit wide) capable of read-write 
from/to an SRAM cell in one clock cycle. 
RMT architecture allows \textit{word packing} to optimize the SRAM usage. 


\textbf{Egress stage}: After finishing the ingress stage processing, 
the packet is submitted to the egress port's queue. 
The queue management process is not P4 programmable, hence we leave its discussion. 
Once the packet is picked from the egress port's queue, it undergoes the egress stage processing. 
The egress stage is similar to the \textit{ingress stage} and shares the same physical components for their processing. 

\textbf{Deparser}: After egress stage processing is finished, the \textit{deparser}
recombines the data from the packet header vector fields and the payload. 
Then the packet is finally propagated through the outgoing channels.

\subsection{V1Model Hardware  Specification Language}\label{HSLSection}
A compiler backend requires information about the available resources of a V1Model switch.
However, the openly available P4C compiler does not provide any interface to model it. 
We developed a JSON format-based hardware specification language (\textit{HSL}) to specify the available resources in the programmable 
components of a V1Model architecture based switch. 
Deatils of the language and an example hardware specification 
of the V1Model switch described in~\cite{bosshart2013forwarding} can be found in our full technical report~\cite{P4CBTechReport}.




\section{P4C Intermediate Representation} \label{IR}
The  P4C~\cite{P4C}  frontend (along with the midend) generates a
\textit{target-independent representation} (IR) of the
P4 program in JSON format. Its major components are following:

$\bullet$\textbf{Header Information}:
This is a list of all the headers (including the packet metadata header) used in the P4 program
along with their member fields and their bitwidth. 

$\bullet$\textbf{Parse Graph}:
The \textit{parse-graph} is a directed acyclic graph representation of the parser state machine 
(parsing logic given in the P4 program).
Every node in the \textit{parse graph} represents a header type, and the edges represent a transition of the parser state machine. 
An edge from node $a$ to node $b$  indicates that after parsing header $a$, 
based on a specific value of one of its member header fields 
$b$ will be parsed.

$\bullet$\textbf{Table Dependency Graph (TDG)}:
Processing logic for \textit{ingress} and \textit{egress} stages is written using imperative programming constructs given by 
the P4  programming language. The P4C frontend converts these logics into a directed acyclic graph 
\textit{Table Dependency Graph} (TDG) containing 
the \textit{logical match-action tables} as nodes and 
their \textit{dependencies} as edge. 
Each node describes the set of fields (header and/or metadata fields) to be matched against the table entries,
the types of matching (exact, prefix, ternary, etc.), and the maximum number of entries to be stored in the memory for this table. 
It also describes the set of actions to be executed based on the match result and the address of the next table to be loaded after 
executing the current table. Every path in the TDG represents a chain of logical MATs.
Four types~\cite{jose2015compiling} of 
dependencies can arise between any two  logical MATs (\textit{A} comes first and then \textit{B}) in the same path. 
These dependencies influence how the logical MATs in a TDG will be mapped on the physical MATs of an RMT pipeline.

\section{Proposed Compiler Backend} \label{MappingProblem}

In this section we present the proposed compiler backend and how it computes the 
header mapping, parse graph mapping and the TDG mapping from the IR of a P4 program. 
While addressing these problems, the compiler backend needs to ensure that a computed mapping does not violate the 
constraints (both architectural and 
resource constraints) of the target hardware and the control flow of the P4 program. 
Besides computing a valid mapping, the backend also needs to maximize the concurrency and resource usage efficiency of the P4 program. 
The compiler backend takes two types of information as input for this purpose: 
a) the hardware resource specification of the switch described using the HSL presented in sec.~\ref{HSLSection}
and b) the hardware-independent intermediate representation (IR) (sec.~\ref{IR}) of the given P4 program generated by the 
P4C~\cite{P4C} compiler frontend. Due to space concerns, fine grained deatils and examples of our compiler backend is not presented in this 
paper. However the full details can be accessed at~\cite{P4CBTechReport}.


\subsection{Header   Mapping} \label{sec:HeaderMapping}
The header fields used in the P4 program is needed to be accommodated using the limited number of fixed bitwidth 
fields in the  PHV.
The problem of optimally allocating the PHV fields (set of items) to accommodate the P4 program's header fields (set of resources) 
can be modeled as a \textit{multiple knapsack} problem~\cite{puchinger2010multidimensional}. 
In this case, we need to find how to optimally fill up a set of bins (the header fields in the P4 program) using the set of items 
(all the PHV fields) while minimizing the waste. 
As the problem is computationally intractable, we used the largest header field first heuristic-based 
algorithm to compute the mapping. 
\subsection{Parse Graph Mapping} \label{sec:ParseGraphMapping}
The compiler backend needs to convert the parsing logic (provided as 
the \textit{parse graph} in the IR~\ref{IR}) into a state table which needs to be accommodatable within the 
dimensions of the parser TCAM. 
Generating an optimal state table from a parse graph and relevant TCAM entries is  a computationally intractable problem~\cite{gibb2013design}.
Our compiler backend uses the heuristic algorithm presented in~\cite{gibb2013design} to find the clusters in the 
\textit{parse-graph} which can be accommodated within the capacity (lookup in the packet header and extracting 
header fields) of the parser hardware. 
For every unique pair of a cluster and an edge (transition in the parser state machine) in the 
cluster graph, one entry is required in the TCAM based parse-table. 
If all the entries can not be accommodated using the parser TCAM, then the P4 program is rejected. 


\subsection{TDG Mapping}\label{sec:TDGMapping}
The TDG expresses the packet processing logic of a P4 program as a control flow among logical 
match-action tables guided by their dependencies. 
The compiler backend needs to map these logical MATs over the physical MATs of the V1Model switch 
while preserving the original control flow and the target hardware resource constraints~\cite{jose2015compiling}. 
Besides this, the compiler backend needs to minimize the  amount of resource usage in each of the used match-action stages. 
The first one leads to reduced processing delay 
of a packet~\cite{jose2015compiling}. The second one leads to an overall reduction in resource usage and power 
consumption by a V1Model switch. 
Finding  a mapping that has optimal resource usage and does not violate the upper mentioned constraints is 
a computationally intractable problem~\cite{jose2015compiling,vass2020compiling}.

\subsubsection{\textbf{Our approach}}\label{OurApproach}
The TDG mapping  problem can be modeled as an 
\textit{integer linear programming}~\cite{jose2015compiling} problem, 
and an optimal mapping can be 
determined (if it exists). However, it may take a large amount of time~\cite{jose2015compiling,vass2020compiling} 
to compute such a mapping. 
Hence we applied heuristic-based algorithms to compute the mapping. The algorithm works in follwoing steps:

\paragraph{Level Generation} \label{Mapping:LevelGeneration}
In the TDG, there can be multiple dependencies between logical MATs in the same path; 
however, the strictest dependency (among the four types dependencies) 
is  the main factor affecting the mapping decision~\cite{jose2015compiling}. 
The mapping algorithm keeps the \textit{strictest} dependency and removes other dependencies. 
Then the TDG nodes are topologically ordered to label them with their appropriate \textit{level}s. 
This ordering ensures that every node with dependency is
assigned a higher \textit{level} than its successors. If two neighbor
nodes in the TDG only have successors or reverse match dependencies
or no dependency, they can be mapped to the same physical match-action stages 
(as they can be executed concurrently or speculatively)~\cite{bosshart2013forwarding,jose2015compiling,sivaraman2016packet} 
and assigned the same \textit{level}.



\paragraph{Mapping Logical Tables}
After \textit{level} generation, all the logical MATs with the same \textit{level}
imply they can be mapped to the same physical match-action stages. 
These logical MATs contains no match, action, or stateful dependencies, and  
these logical MATs can be executed concurrently on the same math-action stage. 
However, a single match-action stage may not contain enough hardware resources to accommodate them. 
In that case, they are mapped to one or multiple consecutive math-action stages. 
All the logical MATs with same level can be ordered and selected for mapping to the physical 
match-action stage using various heuristics~\cite{jose2015compiling}. In this work, 
we prioritized the logical MATs with non-exact (ternary, lpm, or prefix) match fields and mapped them to 
TCAM-based physical match-action blocks. Next, the logical MATs with exact match fields are mapped 
to SRAM-based physical match-action units. These logical MATs can spill into 
TCAMs if it runs out of SRAMs. Ties are broken by at first mapping the logical MAT that appears in the TDG earlier. 



In allocating SRAM blocks for the match, action, and stateful memory entries, the mapping algorithm utilizes the 
\textit{memory packing} feature of RMT architecture. The mapping algorithm 
tries to store multiple entries in a \textit{packing unit} of up to $p_f$  SRAM blocks to reduce SRAM waste and fragmentation. 
The value of $p_f$ is configurable at compile time. Currently, our compiler backend allocates SRAM at block level granularity. 
As a result, an SRAM block is allocated exclusively for the match or action entries of only one MAT; or for accomodating the indirect 
stateful memories. It can lead to a waste of SRAM resources when a small number of entries are required. We leave the goal of 
improving the SRAM utilization as future work. 
The number of action entries required for a logical MAT can be determined at compile time~\cite{robin2022clb} or 
unknown beforehand~\cite{jose2015compiling}.  
Our compiler backend can reserve either a fixed number of action entries or 
one action entry for every MAT entry in a logical MAT.

\section{Implementation and Evaluation}  \label{ImplementationAndEvaluation}
\textbf{Implementation}:
Our P4 compiler backend is implemented in the Python 3 programming language. 
For the frontend, it relies on the P4C~\cite{P4C},
which parses the P4 source code and generates the intermediate representation in JSON format. 
Our backend parses  this IR and stores it in a graph-based data structure. 
For computing the \textit{parse graph} to \textit{state table}
representation for the parser TCAM,
we relied on the algorithms proposed in~\cite{gibb2013design}. 
This project's source code was written in Python 2 language, and it was not designed for P4\textsubscript{16}'s 
intermediate representation of P4C. We converted the source code in 
Python 3 language and integrated it with our project with a moderate level of modifications. 
All source code of the project is publicly available~\cite{P4CB} under an open-source license.







\textbf{Evaluation}:
To the best of our knowledge, there are no openly available complete compiler backend (which can compute all three types of mapping for a P4 program) 
for RMT switches and benchmark p4 programs with their resource consumption 
in the V1Model hardware pipeline. 
To this end, we have selected the following four P4 programs to evaluate  our compiler backend's performance:
\textit{a) IPv4/IPv6 QoS modifier}: a P4 program where a stateful memory is accessed by different P4 tables in different path in the TDG. 
\textit{b) Simple layer-2/3 forwarding}: a P4 program containing all four types of dependecy, 
\textit{c) Complex layer-2/3 forwarding}: complex version of previous one with large SRAM and TCAM requirement and 
\textit{d) Traffic Aanonymizer}: a computation intensive P4 program. 
The first one is selected to show our compiler backend's capability of handle 
complex P4 programs with stateful memory access. Other three programs are selected as they
reported~\cite{kim2019ontas,jose2015compiling} to be realizable using Tofino~\cite{tofino2,opentofino} switches.
For each P4 program, we used the intermediate representation (IR) generated by the P4C frontend and 
the V1Model switch described in ~\cite{bosshart2013forwarding} as the benchmark hardware. 
All experiments were run on  an HP laptop with an Intel Core-i7 processor,
24 GB RAM, running Ubuntu 20.04.

\begin{table}[t]

\centering{
    \small{
      \begin{tabular}{|c|c|c|c|c|}
        \hline
        Program Name  & \begin{tabular}[c]{@{}c@{}}(\#  Header  Fields,\\   Total Bitwidth)\end{tabular} & \begin{tabular}[c]{@{}c@{}}Total \\ PHV \\ Bitwidth\end{tabular} & \begin{tabular}[c]{@{}c@{}}Waste\\ ( \% )\end{tabular} & \begin{tabular}[c]{@{}c@{}}Ex. time \\ (in ms)\end{tabular} \\ \hline
        QoS-Modifier   & (66, 1288)                                                                                                            & 1432                                                             & 10.05                                                    & 2                  \\ \hline
        Traffic-anony & (58, 1064)                                                                                                            & 1208                                                             & 11.92                                                    & 2.03               \\ \hline
        L2L3-simple   & (126, 2912)                                                                                                           & 3088                                                             & 5.69                                                     & 2.12               \\ \hline
        L2L3-Complex  & (94, 1976)                                                                                                            & 2112                                                             & 6.43                                                     & 108                \\ \hline
        \end{tabular}
}
}

\caption{\small{Total number of header fields used in P4 programs and their bitwdiths,
total bitwidth of required PHV fields, percentage of waste in PHV fields and the total execution time 
required for computing the \textit{\textbf{header mapping}}.}}
\label{tab::headercomparison}
\end{table}

\subsubsection{Result Analysis}\label{ResultAnalysis}

\textbf{Header Mapping}: Table~\ref{tab::headercomparison} shows the result of our compiler backend's 
\textit{header mapping} of the benchmark P4 programs.
The small programs (\textit{QoS-Modifier} and \textit{L2L3-Simple}) consume around 30-35\% 
and  the large programs (\textit{Traffic-Anony} and \textit{L2L3-Complex}) 
consume around 51\% of the PHV's capacity. Some space in the 
PHV fields are wasted (approx. 10-12\% and 5.7-6.4\% for small and large programs respectively) 
due to required padding. For all the P4 programs, the time required to compute the mapping is short and ranges 
between 2 to 3 milliseconds (approx.).

\begin{table}[]
\centering
{
    \small{
      \begin{tabular}{|c|c|c|c|c|}
        \hline
        Program Name  & \begin{tabular}[c]{@{}c@{}}\# States \\ in Parse \\ Graph\end{tabular} & \begin{tabular}[c]{@{}c@{}}\# Edges \\ in Parse \\ Graph\end{tabular} & \begin{tabular}[c]{@{}c@{}}Req. \\ TCAM \\ Entries\end{tabular} & \begin{tabular}[c]{@{}c@{}}Ex. time \\ (in ms)\end{tabular} \\ \hline
        QoS-Modifier   & 5                                                                      & 8                                                                     & 10.05                                                           & 31                                                          \\ \hline
        Traffic-anony & 4                                                                      & 10                                                                    & 11.92                                                           & 21                                                          \\ \hline
        L2L3-simple   & 11                                                                     & 31                                                                    & 5.69                                                            & 132                                                         \\ \hline
        L2L3-Complex  & 7                                                                      & 14                                                                    & 6.43                                                            & 65                                                          \\ \hline
        \end{tabular}
}
}
\caption{\small{
  Total number of states and edges in the parse graph of the
  P4 programs, number of TCAM entries  required for the \textit{state table} and 
  the total execution time required for computing the \textit{\textbf{parse graph mapping}}.}}
\label{tab::parserComparison}  
\end{table}

\textbf{Parse Graph Mapping}: 
Table~\ref{tab::parserComparison} shows the result of our compiler backend's 
\textit{parse graph mapping} of the benchmark P4 programs. 
The benchmark hardware contains 256$\times$40b parser TCAM. It looks at 48 bytes into the packet,
identify a maximum of four headers and extract 48 bytes of data in every cycle. 
For the complex P4 program (\textit{L2L3-Complex}) with 11 states and 
31 edges in the parse graph, it requires only 22 entries in the TCAM, which is less than 9\% of the total capacity of the parser TCAM. 
The \textit{parse graph} is simpler for the
rest of the programs and consumes only 2\% of the TCAM's capacity. 
The mapping algorithm's execution time (rightmost column) increases with the complexity (nuumber of nodes and edges) of the 
\textit{parse-graph}. 


\begin{table*}[!t]
\centering{
  \begin{tabular}{|c|c|c|c|c|c|c|c|}
    \hline
    Program Name  & \# Nodes in TDG & \# Edges in TDG & Stages & Latency (in cycle) & \# TCAM Block Usage & \# SRAM Block Usage & Ex. Time (in ms) \\ \hline
    QoS-Modifier   & 16              & 20              & 3      & 38                 & 6                   & 4                   & 31               \\ \hline
    Traffic-anony & 84              & 194             & 18     & 156                & 2                   & 35                  & 1700             \\ \hline
    L2L3-simple   & 24              & 38              & 5      & 42                 & 57                  & 207                 & 41               \\ \hline
    L2L3-Complex  & 60              & 138             & 31     & 108                & 260                 & 995                 & 925              \\ \hline
    \end{tabular}
}
    \caption{\small{Total number of states and edges in the TDG of the
    P4 programs, number of match-action stages required, packet processing latency, total  TCAM blocks required,  SRAM blocks required, and 
    the total execution time required for computing the \textit{\textbf{parse graph mapping}}.}}
\label{tab::TDGmapperComparison} 
\end{table*}

\textbf{TDG Mapping}:
Table~\ref{tab::TDGmapperComparison}
shows the result of our compiler backend's 
\textit{TDG  mapping} of the benchmark P4 programs.
The \textit{QoS-Modifier} program is complex and need \textit{stateful memory} access by different logical MATs in different 
paths in the TDG. 
Our compiler backend  generates a valid mapping for this P4 program. It maps the logical MATs over three physical match-action stages 
 on 6 TCAM blocks and 4 SRAM blocks, and the processing 
latency of every packet under this mapping is 38 cycles.  
For the \textit{Traffic-Anony} program, our compiler backend uses 18 physical match-action stages.
It needs 2 TCAM blocks and 35 SRAM blocks; and the processing 
latency of every packet under this mapping is 156 cycles.  
In the case of \textit{L2L3-Simple} 
our compiler backend uses 5 match-action stages and achieves the packet processing latency of 42 cycles. 
In the case of \textit{L2L3-Complex}, our compiler backend uses 31 stages  and achieves a  packet processing latency of 108 cycles. 
Our compiler backend  allocates SRAM blocks for action memory and indirect stateful memories 
at a granularity of blocks. As a result, when 
the SRAM requirement of two or more logical MATs can be fulfilled using only one SRAM block, our compiler backend 
allocates at least one SRAM block to every logical MAT. This memory packing is less efficient compared to~\cite{jose2015compiling}. 
 We leave the goal of improving the memory packing as a 
future work. 

\textit{Execution time}: In the case of \textit{L2L3-Simple} and \textit{L2L3-Complex}, our compiler backend 
requires 41 and 925 ms respectively. Here the complex version of the program (\textit{L2L3-Complex}) requires 
approximately 20x time  compared to its simpler version ()\textit{L2L3-Simple}).
Our compiler backend needs 31 ms for the {QoS-Modifier} and 1700 ms for the \textit{Traffic-Anony}  program. 
Hence it is clear that with the increase in size and compplexity of the TDG (indicated by increased number of nodes and 
edges in the TDG) our compiler backend needs for execution time.

\section{Discussion}  \label{Discussion}

\textbf{Limitations}:
Our compiler backend supports most of the P4 language constructs covering a wide range of use cases. 
However, it still does not support variable-length header parsing and direct stateful memory access in actions. 
Both of them can be avoided through careful design of the p4 program. 
Besides this, it does not support the atomic transaction mechanism available in the P4 language. 
We are working on supporting these  P4 language features.

\textbf{Extending  V1Model Architecture}:
The PSA~\cite{PSA} or Tofino~\cite{tofino2} is an extension of V1Model architecture where the architectures support different externs. 
These architectures can combine multiple simpler instructions in one atomic instruction to achieve complex functionalities. 
For example, the \textit{register extern} available in Tofino switches~\cite{opentofino} can execute 
four-way branching instructions. It can execute two if-else pairs and read-modify-write operation on a pair of 
registers (indirect stateful memory) using only one extern. 
However, to use them (or any new extern in general) in a P4 program, the P4C compiler frontend needs to support them. 
After that, these externs can be 
supported in our compiler backend with minor modifications to compute mapping for that P4 program. 

\textbf{Writing New Mapping Algorithm}:
Our compiler backend is designed in a modular way. After parsing the intermediate representation of a P4 program, 
it stores the preprocessed information (header information, parse graph, TDG) in various convenient data 
structures (hash table, graph, etc.). 
Besides this, it also stores the resources in a  V1Model  switch in various convenient data structures 
(hashtable, array, etc.). As an open-source project, researchers can reuse this processed information to 
write newer algorithms for header mapping, parse graph mapping, and TDG mapping. 
Detail discussion on source code organization is available in~\cite{P4CB}.




\section{Conclusion}  \label{Conclusion}
We have presented an open-source compiler backend that can map a P4 (version 16) program to the hardware resources 
of a  V1Model switch. 
It uses heuristic-based algorithms to compute this mapping and give a quick decision on the realizability of a P4 program. 
We believe this open-source compiler backend can serve as a cost-effective platform for analyzing the realizability 
and resource consumption 
of a P4 (version 16) program in real-world V1Model  switches. 
While it will be challenging to extend the backend for radically different packet processing hardware architecture, 
but derivative of V1Model may not be out of reach (as discussed in sec.~\ref{Discussion}). As we speak, the P4 community 
is already working toward the \textit{portable-switch-architecture}. An extensible compiler can 
help network researchers accelerate the complete cycle of innovation on these architectures.
Our paper contributes a complete compiler backend for P4\textsubscript{16} programs and its proof of concept. 
This can provide 
an open platform to the programmable switch researchers for experimenting with different mapping algorithms and different 
derivatives of the V1Model switch. Overall, it can help the network researchers, and 
the conversation can begin for 
accelerating the full cycle innovation in the programmable switch domain.

\bibliographystyle{unsrt}
\bibliography{Compiler}

\end{document}